\documentclass[12pt]{article}
\usepackage{epsf,amsmath,amssymb,amsfonts}
\usepackage{graphicx}
\usepackage{epsfig}
\usepackage{verbatim}
\usepackage{cite}
\newcommand{\be}{\begin{equation}}
\newcommand{\ee}{\end{equation}}
\newcommand{\bea}{\begin{eqnarray}}
\newcommand{\eea}{\end{eqnarray}}
\newcommand{\ba}{\begin{eqnarray}}
\newcommand{\ea}{\end{eqnarray}}

\newcommand{\beq}{\begin{equation}}
\newcommand{\eeq}{\end{equation}}
\newcommand{\beqa}{\begin{eqnarray}}
\newcommand{\eeqa}{\end{eqnarray}}
\newcommand{\beqar}{\begin{eqnarray*}}
\newcommand{\eeqar}{\end{eqnarray*}}
\numberwithin{equation}{section}

\begin{document}
\setlength{\unitlength}{1mm}

\thispagestyle{empty} \rightline{\hfill \small hep-th/0607222
 } \rightline{\hfill \small
 ITP-UU-06/32  }
\rightline{\hfill \small  SPIN-06/28 } \rightline{\hfill \small
 ITFA-2006-26 }

\vspace*{2cm}

\begin{center}
{\bf \Large The Gravitational Description of Coarse Grained
Microstates}\\

\vspace*{1.4cm}

\vspace*{0.3cm}

{\bf Luis F. Alday,}$^1\,$ {\bf Jan de Boer,}$^2\,$ {\bf Ilies
Messamah}$^2$

\vspace{.8cm}

{\it $^1\,$Institute for Theoretical Physics and Spinoza Institute, Utrecht University \\ 3508 TD Utrecht, The Netherlands}\\[.3em]
{\it $^2\,$Instituut voor Theoretische Fysica,\\ Valckenierstraat
65, 1018XE Amsterdam, The Netherlands.}

\vspace*{0.2cm} {\tt l.f.alday@phys.uu.nl, jdeboer@science.uva.nl,
imessama@science.uva.nl}

\vspace{.8cm}
{\bf Abstract}
\end{center}

In this paper we construct a detailed map from pure and mixed
half-BPS states of the D1-D5 system to half-BPS solutions of type
IIB supergravity. Using this map, we can see how gravity arises
through coarse graining microstates, and we can explicitly confirm
the microscopic description of conical defect metrics, the $M=0$
BTZ black hole and of small black rings. We find that the entropy
associated to the natural geometric stretched horizon typically
exceeds that of the mixed state from which the geometry was
obtained.

\noindent

\vfill \setcounter{page}{0} \setcounter{footnote}{0}
\newpage

\tableofcontents

\newpage

\setcounter{equation}{0}

\section{Introduction}

The AdS/CFT correspondence and the counting of microstates of
black holes in string theory provide a substantial amount of
evidence that gravity is thermodynamic in nature, and that
classical gravity arises by coarse graining over a (large number
of) microstates. This point of view was elaborated in
\cite{Balasubramanian:2005mg} for large black holes in AdS${}_5$
and also for 1/2 BPS geometries that asymptote to AdS${}_5\times
S^5$, following the work of \cite{Lin:2004nb} (for related work
see also
\cite{e0,e1,Buchel:2004mc,{Suryanarayana:2004ig},{Shepard:2005zc},{Silva:2005fa},e2,e3,e4,e5}).
In the 1/2 BPS case one can develop a precise map between
microstates and geometries and study the coarse graining in
detail. The Hilbert space is that of $N$ free fermions in an
harmonic oscillator potential, and given a state or ensemble in
that Hilbert space one can associate a classical phase space
density to it. This phase space density, a function on a
two-dimensional plane, then completely determines the
ten-dimensional metric. Using this setup one finds, as expected,
that almost all states are typical, i.e. very similar to the
ensemble average, and very difficult to distinguish from each
other.

The purpose of this paper is to generalize and study the map
between 1/2 BPS states and geometries to the AdS${}_3$ case. Many
aspects of this system were previously studied in
\cite{Balasubramanian:2000rt,Maldacena:2000dr,Lunin:2001jy,Lunin:2002qf,Lunin:2002bj,Lunin:2002iz,Bena:2004tk,Mathur:2005zp,Iizuka:2005uv,Taylor:2005db,Balasubramanian:2005qu,Alday:2005xj,jevicki,Rychkov:2005ji}.
The AdS${}_3$ case is interesting, as it has a somewhat richer
structure, harbors many well-known solutions such as conical
defect metrics, the BTZ black hole and black rings, and plays an
important role in most of the microscopic derivations of black
hole entropy. It is also the context in which Mathur formulated
his fuzzball picture of black holes (see e.g.
\cite{Mathur:2005zp}). Though there are no macroscopic 1/2 BPS
black holes in AdS${}_3$ we will encounter several black-hole like
features in our description. As we will show, the technical
details are quite different from the AdS${}_5$ case.

The relevant 1/2-BPS geometries were obtained by dualizing
solutions describing classical string profiles in
\cite{Lunin:2001jy}. The classical string profile corresponds to a
certain parametrized curve $\mathbf{F}(s)\subset \mathbb R^4$.
This is not yet the most general solution, as the string can also
oscillate in four other transversal directions, which we take to
be a four-torus, and in addition has fermionic excitations
(studied in \cite{Taylor:2005db}). However, in our paper, we will
not consider these additional degrees of freedom.

The classical phase space of gravitational solutions is the set of
curves $\mathbf{F}(s)$ of fixed ``length'' $N\sim \int ds |
\dot{\mathbf{F}}(s)|^2$. The symplectic form on the phase space
was derived in \cite{jevicki,Rychkov:2005ji}. From this one infers
that the Fourier modes of $\mathbf{F}$ correspond to standard free
bosonic string oscillators without the zero mode, with the length
corresponding to the energy or $L_0$ eigenvalue of a state. The
Hilbert space in question is therefore the set of states of level
$N$ in the Hilbert space of four free bosons. To a state or
density matrix in this Hilbert space we will associate a phase
space density which is a measure on the space of loops of fixed
length. This measure will then be used to construct the explicit
1/2 BPS metric. Quantizing a subset of the degrees of freedom of
the metric is a familiar procedure as all minisuperspace
approaches to quantum gravity use exactly the same idea. The usual
complaint about minisuperspace approximations is that the
approximation is not controlled, i.e. it is not the leading term
of the expansion in some small parameter. The same complaint in
principle also applies to our construction, though supersymmetry
will make the results more robust, and we believe that in view of
the similarity with e.g. the discussion of large black holes in
AdS${}_5$ in \cite{Balasubramanian:2005mg} we are still learning
valuable lessons about quantum gravity despite the 1/2-BPS
restriction.

This paper is organized as follows. The details of the map between
microstates and geometries will be discussed in section~2.
Coherent states will play an important role, and we will also
discuss some subtleties associated to the choice of phase space
density. Furthermore, we consider in detail the case of a circular
profile, showing that for large quantum numbers the geometry given
by our map differs by a small correction from that corresponding
to a classical circular curve.

In section~3 we will show how we can use our setup to construct a
class of generalized conical defect metrics, which include conical
defect metrics with deficit angle $2\pi/n$ with $n$ integer. We
will also confirm the claim of \cite{Lunin:2002iz} that metrics
with deficit angle $\alpha=2\pi/n$ with $n$ not an integer cannot
be constructed in this way, and our new metrics are the best
approximation to conical deficit metrics with such values of
$\alpha$.

In section~4 we study the metrics associated with various
ensembles, in particular for the $M=0$ BTZ black hole, small black
rings and generic thermal ensembles in absence of a condensate.
For the $M=0$ BTZ black hole we will find that the area of the
surface (the `stretched horizon') within which the metric differs
significantly from that of the $M=0$ BTZ black hole yields an
entropy proportional to $N^{3/4}$, whereas the logarithm of the
number of states scales as $N^{1/2}$. Thus, typical states are
larger than what one would expect from a naive fuzzball picture of
black holes, and we will discuss possible implications of this
observation.

The ensembles we study are all of the form $\rho \sim \exp(-\sum_i
a_i O_i)$. For these, the entropy obeys \be dS = \sum_i a_i
d\langle O_i \rangle \ee and such ensembles are therefore the most
natural candidate dual descriptions of black objects that obey the
first law. We will indeed find a rather universal behavior in the
ensembles that we study, including the ensemble that we proposed
as dual description of the small black ring in
\cite{Alday:2005xj}.

Some concluding remarks and open problems are finally given in
section~5.

\section{The map betweeen states and geometries}

\subsection{Conventions}

We will follow the conventions of \cite{Rychkov:2005ji}. By
dualizing a fundamental string with transversal profile
$\mathbf{F}(s)\subset\mathbb R^4$ we obtain the following
microstate geometries of the $D1-D5$ system, written in string
frame
\begin{eqnarray}
\label{genericsol} ds^2=\frac{1}{\sqrt{f_1
f_5}}\left[-(dt+A)^2+(dy+B)^2 \right]+\sqrt{f_1 f_5}
d\mathbf{x}^2+\sqrt{f_1/f_5}d
\mathbf{z}^2 \nonumber\\
e^{2 \Phi}=\frac{f_1}{f_5},\hspace{0.2in}C=\frac{1}{f_1}\left(dt+A
\right)\wedge \left( dy+B \right)+\mathcal{C} \nonumber \\
dB=*_4 dA,\hspace{0.2in}d\mathcal{C}=-*_4 df_5 \label{aux1} \\
f_5=1+\frac{Q_5}{L}\int_0^L
\frac{ds}{|\mathbf{x}-\mathbf{F}(s)|^2}\\
f_1=1+\frac{Q_5}{L}\int_0^L \frac{|\mathbf{F}'(s)|^2
ds}{|\mathbf{x}-\mathbf{F}(s)|^2} \nonumber\\
A=\frac{Q_5}{L} \int_0^L \frac{F'_i(s)
ds}{|\mathbf{x}-\mathbf{F}(s)|^2}
\end{eqnarray}
The solutions are asymptotically $\mathbb R^{1,4} \times S^1
\times T^4$, $y$ parametrizes the $S^1$ which has coordinate
radius $R$, and $\mathbf{z}$ are coordinates on the $T^4$ which
has coordinate volume $V_4$. The Hodge duals $*_4$ are defined
with respect to the four non-compact transversal coordinates
$\mathbf{x}$. We can take a decoupling limit which simply amounts
to erasing the $1$ from the harmonic functions. The resulting
metric will then be asymptotically equal to AdS${}_3 \times S^3
\times T^4$.

As mentioned above, the solutions are parametrized in terms of a
closed curve
\begin{equation}
x_i=F_i(s),\hspace{0.2in}0<s<L,~i=1,...,4.
\end{equation}
and we will ignore oscillations in the $T^4$ direction as well as
fermionic excitations in this paper.

The number of $D1$ and $D5$ branes is denoted by $N_1$ and $N_5$,
and they are related to the charges $Q_i$ by
$$Q_5=g_s N_5 \hspace{0.2in} Q_1=\frac{g_s}{V_4} N_1$$
The parameter $L$ has to satisfy
\begin{equation}
L=\frac{2\pi Q_5}{R} .
\end{equation}
Besides, the curve has to satisfy the following relation
\begin{equation} \label{length}
Q_1=\frac{Q_5}{L} \int_0^L |\mathbf{F}'(s)|^2 ds
\end{equation}
which reflects the fact that the original string had a fixed
length. It turns out that the space of classical solutions has
finite volume and therefore will yield a finite number of quantum
states. Indeed, expanding $\mathbf{F}$ in oscillators: \be
\mathbf{F}(s)=\mu \sum_{k=1}^\infty
\frac{1}{\sqrt{2k}}\left(\mathbf{c}_k e^{i \frac{2\pi k
}{L}s}+\mathbf{c}_k^\dagger e^{-i \frac{2\pi k }{L}s} \right) \ee
where \be \mu=\frac{g_s}{R \sqrt{V_4}} \ee
it was first shown in
\cite{jevicki} (see also \cite{Rychkov:2005ji}) by computing the
restriction of the Poisson bracket to the space of solutions
(\ref{aux1}) that
\begin{eqnarray}
~[c_k^i,c_{k'}^{j \dagger}]=\delta^{ij}\delta_{kk'}\label{comrel}\\
\left< \int_0^L :|\mathbf{F}'(s)|^2: ds\right>=\frac{(2\pi)^2}{L}~
\mu^2 N  \label{qlength} \\
N\equiv N_1 N_5 =\sum_{k=1}^{\infty} k \left< \mathbf{c}^\dagger_k
\mathbf{c}_k\right> .
\end{eqnarray}
Clearly, the number of states is finite. Using the above quantum
mechanical system, we can now go ahead and construct a map between
the quantum states of the theory and classical field
configurations. As familiar from quantum mechanics, this map will
involve the phase space distribution associated to quantum states.

\subsection{Proposal for the map}

\label{sect22}

Chiral primary operators in the dual CFT are in one to one
correspondence with the states at level $N$ of a Fock space built
out of 8 bosonic and 8 fermionic oscillators (or 24 bosonic
oscillators if we replace $T^4$ by $K3$). Since we are only
interested in fluctuations in the transverse $\mathbb R^4$ we will
keep only four of the bosons and discard the fermions. The Hilbert
space is thus spanned by
\begin{equation} \label{hilbert}
|\psi \rangle=\prod_{i=1}^4 \prod_{k=1}^\infty(c^{i \dagger}
_{k})^{N_{k_i}}|0\rangle,\hspace{0.3in}\sum k N_{k_i}=N
\end{equation}
Given a state, or more generically a density matrix in the CFT
\begin{equation} \label{density}
\rho=\sum_i | \psi_i \rangle \langle \psi_i |
\end{equation}
we wish to associate to it a density on phase space. The phase
space is given by classical curves which we will parametrize as
(note that $d$ and $\bar{d}$ are now complex numbers, not
operators)
\begin{equation} \label{classcurv}
\mathbf{F}(s)=\mu \sum_{k=1}^\infty
\frac{1}{\sqrt{2k}}\left(\mathbf{d}_k e^{i \frac{2\pi k
}{L}s}+\bar{\mathbf{d}}_k e^{-i \frac{2\pi k }{L}s} \right)
\end{equation}
and which obey the classical constraint (\ref{length}).

We now propose to associate to a density matrix  of the form
(\ref{density}) a phase space density of the form
\begin{equation} \label{proposal}
f(d,\bar{d})= \sum_i \frac{\langle 0|
e^{\mathbf{d}_k\mathbf{c}_k}|\psi_i \rangle \langle \psi_i |
e^{\mathbf{\bar d}_k\mathbf{c}_k^\dagger}|0 \rangle }{\langle 0|
e^{\mathbf{d}_k\mathbf{c}_k} e^{\mathbf{\bar
d}_k\mathbf{c}_k^\dagger}|0 \rangle } .
\end{equation}
Notice that this phase space density, as written, is a function on
a somewhat larger phase space as $d,\bar{d}$ do not have to obey
(\ref{length}). We will discuss this issue in the next section and
ignore it for now.

The density (\ref{proposal}) has the property that for any
function $g(d,\bar{d})$ \be \label{prop} \int \int_{d,\bar{d}}
f(d,\bar{d}) g(d,\bar{d}) = \sum_i \langle \psi_i |
:g(c,c^{\dagger}):_A | \psi_i \rangle \ee where
$:g(c,c^{\dagger}):_A$ is the anti-normal ordered operator
associated to $g(c,c^{\dagger})$, and $\int_{d,\bar{d}}$ is an
integral over all variables $d_i$. It is possible to construct
other phase densities such as the Wigner measure where anti-normal
ordering is replaced by Weyl ordering, or one where anti-normal
ordering is replaced by normal ordering. Though apparently
different, they will yield identical results if we are interested
in computing expectation values of normal ordered operators. Since
the theory behaves like a $1+1$ dimensional field theory this is
certainly the natural thing to do in order to avoid infinite
normal ordering contributions. Besides, everything we do is
limited by the fact that our analysis is in classical gravity and
therefore can at best be valid up to quantum corrections.

To further motivate (\ref{proposal}) we notice that it associates
to a coherent state a density which is a gaussian centered around
a classical curve, in perfect agreement with the usual philosophy
that coherent states are the most classical states. It is then
also clear that given a classical curve (\ref{classcurv}) we wish
to associate to it the density matrix
\be \label{classdens}
\rho = P_N e^{\mathbf{d}_k\mathbf{c}_k}|0 \rangle \langle 0 |
e^{\mathbf{\bar d}_k\mathbf{c}_k^\dagger} P_N
\ee
where $P_N$ is the projector onto the actual Hilbert space of
states of energy $N$ as defined in (\ref{hilbert}). Because of
this projector, the phase space density associated to a classical
curve is not exactly a gaussian centered around the classical
curve but there are some corrections due to the finite $N$
projections. Obviously, these corrections will vanish in the
$N\rightarrow \infty$ limit.

Since the harmonic functions appearing in (\ref{aux1}) can be
arbitrarily superposed, we finally propose to associate to
(\ref{density}) the geometry
\begin{eqnarray} \label{finalprop}
f_5 & = & 1+\frac{Q_5}{L} {\cal N}\int_0^L \int_{d,\bar d} \frac{
f(d,\bar{d})
ds}{|\mathbf{x}-\mathbf{F}(s)|^2} \nonumber \\
f_1 & = & 1+\frac{Q_5}{L}{\cal N}\int_0^L \int_{d,\bar d} \frac{
f(d,\bar{d}) |\mathbf{F}'(s)|^2
ds}{|\mathbf{x}-\mathbf{F}(s)|^2} \nonumber \\
A^i & = & \frac{Q_5}{L}{\cal N}\int_0^L \int_{d,\bar d} \frac{
f(d,\bar{d})\mathbf{F}'_i(s) ds}{|\mathbf{x}-\mathbf{F}(s)|^2}
\end{eqnarray}
with the normalization constant
\begin{equation}
{\cal N}^{-1}=\int_{d, \bar d} f(d,\bar d)
\end{equation}

It is interesting to contrast this approach to the results of
\cite{Balasubramanian:2005qu}. In that paper, properties of the
geometry were derived from microstates by evaluating two-point
functions in the CFT. Assuming that the two-point functions do not
renormalize from weak to strong coupling, this provides a direct
probe of the geometry, but it is not easy to reconstruct the
geometry directly. Despite this, one can see very nicely that
coarse graining leads to classical gravitational descriptions, in
accordance with our findings.

In \cite{Lunin:2002iz} it was shown that the geometries
corresponding to a classical curve are regular provided
$|\mathbf{F}'(s)|$ is different from 0 and the curve is not self
intersecting. In our setup we sum over continuous families of
curves which generically smoothes the singularities. The price
that one pays for this is that the solutions will no longer solve
the vacuum type IIB equations of motion, instead a small source
will appear on the right hand side of the equations. Since these
sources are subleading in the $1/N$ expansion and vanish in the
classical limit, they are in a regime where classical gravity is
not valid and they may well be cancelled by higher order
contributions to the equations of motion.

The distribution corresponding to a generic state $|\psi \rangle=
\prod_{k=1}^\infty(c^{i \dagger}_{k})^{N_{k_i}}|0\rangle$ can be
easily computed
\begin{equation}
\label{genericwigner} f(d,\bar d)=\prod_{k,i} (d_k^i
\bar{d}_k^i)^{N_{k_i}}e^{-d_k^i \bar{d}_k^i}
\end{equation}
As a check, we will verify that (\ref{qlength}) is satisfied. To
do so, we need to come up with an operator which reproduces the
left-hand side of (\ref{qlength}) upon anti-normal ordering. In
view of (\ref{comrel}), this is relatively easy to implement, for
example in $|\mathbf{F}'(s)|^2$ we simply need to replace \be
\label{aux11} d^i_k \bar{d}^i_k \rightarrow d^i_k \bar{d}^i_k-1.
\ee We will continue to write expressions like
$|\mathbf{F}'(s)|^2$ in order to not clutter the notation, but
always keep in mind that shifts like (\ref{aux11}) may be
necessary in order to keep track of the proper normal ordering of
the operator in question. Using (\ref{aux11}) it is then easy to
show that (\ref{qlength}) is satisfied. Indeed, (\ref{aux11}) is
equivalent to the following condition
\begin{equation}
Q_1=\frac{Q_5}{L} {\cal N} \int_0^L \int_{d,\bar d} f(d,\bar{d})
|\mathbf{F}'(s)|^2 ds
\end{equation}
and this is satisfied as a consequence of $\sum k N_{k_i}=N_1
N_5$. More explicitly
\begin{eqnarray} \nonumber
\label{level} \frac{Q_5}{L} {\cal N} \int_0^L \int_{d,\bar d}
f(d,\bar{d}) |\mathbf{F}'(s)|^2 ds = \frac{Q_5}{L} {\cal N}
\int_{d,\bar d} f(d,\bar{d}) \left( \mu^2 \frac{4 \pi^2}{L^2} L
\sum_{k=1}^\infty
k (d^i_k \bar{d}_k^i-1) \right) =\\
=\mu^2 \frac{4 \pi^2}{L^2} Q_5 \left( \sum_{k} k N_{k_i}
\right)=Q_1.
\end{eqnarray}
To go from the first line to the second we have used the following
identity
\begin{equation}
\int_{d,\bar d} (d \bar{d})^k e^{-d\bar{d}}=4 \pi \int_0^\infty dr
r^{2k+1}e^{-r^2}=2 \pi k!.
\end{equation}

\subsection{Reparametrization invariance and microcanonical vs canonical}

There is an important subtlety that we need to address. We wish to
study the phase space of curves of fixed length. The phase space
of curves of arbitrary length is very easy, it simply consists of
an infinite set of harmonic oscillators. The length of the curve
is measured by some operator $\hat{N}$. The constraint $\hat{N}=N$
is however first class in the language of Dirac, because
$[\hat{N},\hat{N}]=0$ (or in classical language, the length
Poisson commutes with itself). First class constraints generate a
gauge invariance. In the present case, the operator $\hat{N}$ also
generates a gauge invariance, which is simply the shift of the
parametrization of the curve, \be {\mathbf F}(s)\rightarrow
{\mathbf F}(s+\delta s). \ee This follows immediately from the
commutation relations of $\hat{N}$ with the oscillators.

Therefore, we have two possibilities: we can either not impose the
length constraint, and include an extra factor $\exp(-\beta
\hat{N})$ in the calculations, where we choose $\beta$ such that
the expectation value of $\hat{N}$ is precisely $N$. This would be
like doing a canonical ensemble, and for many purposes this is
probably a very good approximation.

If we insist on fixing the length however, we also have to take
the gauge invariance into account. Therefore, once we include the
length constraint, it is impossible to distinguish curves whose
parametrization is shifted by a constant. In particular, the
expectation value of ${\mathbf F}(s)$ will always be zero, because
the only meaningful quantities to compute are those of gauge
invariant operators, and ${\mathbf F}(s)$ is not gauge invariant.
Notice that $f_1$, $f_5$ and $A$ are gauge invariant so for those
it is not a problem.

We also need to improve the map we discussed above a little bit:
we need to project the measure (\ref{proposal}) on loop space onto
the submanifold of phase space of curves of fixed length. It is
not completely trivial to determine the right measure. To get an
idea we will do the simple example of two oscillators.

We consider ${\mathbb C}^2$ with the usual measure. We wish to
restrict to the submanifold $N=a_1 |z_1|^2 + a_2 |z_2|^2$, and we
wish to gauge fix the $U(1)$ symmetry that maps $z_k \rightarrow
e^{i \epsilon a_k} z_k$. What is the measure that we should use?
In general, if we have a three-manifold with a $U(1)$ action, and
we gauge fix this $U(1)$ the measure on the gauge-fixed
two-manifold is simply the induced measure as long as the $U(1)$
orbits are normal to the gauge fixed two-manifold. So if we
integrate a gauge-invariant operator over the gauge fixed
two-manifold, this is the same as integrating it over the entire
three-manifold, but dividing by the length of the $U(1)$ orbit
through each point. Call the length of this orbit at the point $P$
$\ell(P)$. On the three-manifold (given by $N=a_1 |z_1|^2 + a_2
|z_2|^2$) we have the induced measure. If we call this equation
$f=0$, then the induced measure on the three-manifold is $d^4 x
\delta(f) |df|$, with $|df|$ the norm of the differential $df$. So
all in all we can write the integral of a gauge invariant quantity
$A$ on the two-dimensional submanifold as \be \int d^4 x A(x)
\frac{\delta(f) |df|}{\ell(P)}. \ee The length of the $U(1)$ orbit
is rather tricky, for general $a_1,a_2$ the orbits do not even
close. So we will assume that these numbers are integers. Then up
to an overall constant that depends only on $a_i$ the length of
the orbit is almost everywhere \be \ell(P)=\sqrt{\sum a_i^2
|z_i|^2} \ee with some pathologies if some of the $z_i$ vanish.

Interestingly enough, we now see that $|df|$ and $\ell(P)$ cancel
each other. Thus the only modification in the measure will be to
include an extra delta function of the form \be \label{delta}
\delta(N-\sum_k k d_k \bar{d}_k) \ee in phase space density. As
long as we integrate gauge invariant quantities this will yield
the right answer. Thus, in (\ref{proposal}) and in
(\ref{genericwigner}) we should include the appropriate delta
function.

Inserting the delta function is just like passing from a canonical
to a microcanonical ensemble. For many purposes the difference
between the two is very small, and not relevant as long as we
consider the classical gravitational equations of motion only. We
will therefore in the remainder predominantly work in the
canonical picture, commenting on the difference with the (more
precise) microcanonical picture when necessary.

\subsection{An example: the circular profile}

In the following we will consider the instructive example of a
circular profile. First we will compute the geometry due to a
classical circular curve and then compare the result with the
geometry obtained following the prescription in
section~\ref{sect22}. This will effectively correspond to a
slightly smeared circular profile.

\subsubsection{Classical profile}

\label{sec241}

We consider the following profile
\begin{equation}
\label{circprofile} F^1(s)=a \cos{\frac{2\pi
k}{L}s},\hspace{0.2in}F^2(s)=a \sin{\frac{2\pi
k}{L}s},\hspace{0.2in}F^3(s)=F^4(s)=0
\end{equation}
which describes a circular curve winding $k$ times around the
origin in the $12$-plane. In order to simplify our discussion, we
focus on the simplest harmonic function $f_5$. Plugging
(\ref{circprofile}) into (\ref{genericsol}) it is straightforward
to compute
\begin{equation}
f_5=1+\frac{Q_5}{\sqrt{(x_1^2+x_2^2+x_3^2+x_4^2+a^2)^2-4a^2(x_1^2+x_2^2)}}
\end{equation}
where the value of $a$ is fixed by the condition (\ref{length})
$$Q_1=Q_5 \left(\frac{2 \pi k}{L} a\right)^2.$$
In order to evaluate the various integrals it will be convenient
to Fourier transform the $x$-dependence. Using
\begin{equation}\label{fourier}\frac{1}{|\mathbf{x}|^2}=\frac{1}{4\pi^2}\int d^4 u
\frac{e^{i \mathbf{u}.\mathbf{x}}}{|\mathbf{u}|^2}\end{equation}
we can write $f_5$ in the following equivalent way
\begin{eqnarray} \label{formalexp}f_5^{clas}=1+\frac{Q_5}{4 \pi^2}\int d^4 u \frac{e^{i
\mathbf{u}.\mathbf{x}}}{|\mathbf{u}|^2}J_0(a\sqrt{u_1^2+u_2^2})=\\
=1+J_0 \left( a \sqrt{-\partial_1^2-\partial_2^2}\right)
\frac{Q_5}{|\mathbf{x}|^2}
\end{eqnarray}
Writing $f_5$ in this somewhat formal way has the advantage that
it can be more easily compared to the quantum expression obtained
in the next section. As we will explain in section~\ref{m0btz},
the other harmonic functions can be obtained from the ``generating
harmonic function''
\begin{equation}
\label{besselongeom} f_v=1+Q_5 J_0 \left( a \sqrt{\left( \frac{2
\pi k}{L}v_2+i
\partial_1 \right)^2+\left( \frac{2 \pi k}{L}v_1-i
\partial_2\right)^2} \right) \frac{1}{|\mathbf{x}|^2} .
\end{equation}
For example, putting $v_1=v_2=0$ immediately reproduces
(\ref{formalexp}). The geometry can be written in a more familiar
form by performing the following change of coordinates
\begin{eqnarray} \label{delcoord}
  x_1 = (r^2 + a^2)^{1/2} \sin \theta \cos \varphi \; , \;\;\;\;\; x_2 = (r^2 + a^2)^{1/2} \sin \theta \sin \varphi \nonumber\\
  x_3 = r \cos \theta \cos \psi \; , \;\;\;\;\; x_4 = r \cos \theta \sin
  \psi.
\end{eqnarray}
In terms of these coordinates, the harmonic functions $f_{1,5}$
become in the near horizon limit (i.e. dropping the one)
\begin{equation}
  f_5 =f_v|_{v=0}= \frac{Q_5}{r^2 + a^2 \cos^2 \theta} \; , \;\;\;\;\; f_1=\partial_{v^i} \partial_{v^i} f_v|_{v=0}= \frac{Q_1}{r^2 + a^2 \cos^2 \theta}
\end{equation}
As a consistency check, we notice that $\Box f_5$ is a delta
function with a source at the location of the classical curve, to
be precise $\Box |\mathbf{x}-\mathbf{F}(s)|^{-2} = - 4\pi^2
\delta^{(4)}(\mathbf{x}-\mathbf{F}(s))$. One indeed finds
\begin{eqnarray}
\Box f_5 & = &  -\frac{Q_5}{4 \pi^2 L} \int_0^{L} ds \int d^4 u
e^{i
\mathbf{u}.(\mathbf{x}-\mathbf{F}(s))}=\\
& = & -\frac{Q_5 4 \pi^2}{L} \int_0^{L} ds \delta(x_1-a
\cos\frac{2\pi k }{L} s) \delta(x_2-a \sin\frac{2\pi k }{L} s)
\delta(x_3)
\delta(x_4) =\\
& = & -4 \pi^2 Q_5 \delta(x_1^2+x_2^2-a^2)\delta(x_3)\delta_(x_4)
\label{supp}.
\end{eqnarray}

\subsubsection{Quantum profile}

In a quantum theory it is impossible to localize wave packets
arbitrarily precisely in phase space. Therefore in the quantum
theory we expect to obtain a profile that is something like a
minimal uncertainty Gaussian distribution spread around the
classical curve. If we take the classical circular curve
(\ref{circprofile}) then we associate to it the density matrix
(\ref{classdens}) and subsequently the phase space density
(\ref{proposal}). Working this out we find out that
\begin{equation}
\label{qcircprofile} f(d,\bar{d})=((d_k^1+id_k^2)(\bar{d}_k^1-i
\bar{d}_k^2))^{N/k}e^{-\sum_{l,i} d_l^i \bar{d}_l^i}.
\end{equation}
We have ignored the delta function (\ref{delta}) here and expect
(\ref{qcircprofile}) to be valid for large values of $N/k$. It is
therefore better thought of as a semiclassical profile rather than
the full quantum profile.

According to (\ref{finalprop}) the harmonic function $f_5$ is now
given by
\begin{equation}
f_5=1+\frac{Q_5}{4 \pi^2}{\cal N} \int_0^L ds \int_{d,\bar{d}}
f(d,\bar{d}) \int d^4u \frac{1}{|\mathbf{u}|^2}e^{i
\mathbf{u}.(\mathbf{x}-\mathbf{F}(s)) + \sum_l \frac{u^2
\mu^2}{2l}}
\end{equation}
where we have used (\ref{fourier}) and the constant $\sum_l
\frac{u^2 \mu^2}{2l}$ appears due to the fact that we want to
compute a normal ordered quantity instead of an anti-normal
ordered one. The function $\mathbf{F}(s)$ depends on an infinite
set of complex oscillators $d_{\tilde{l}}^i$. It can be easily
seen that the contribution for the oscillators different from
$d_k^1$ and $d_k^2$ cancels exactly against the normal ordering
constant $u^2\mu^2/2l$ mentioned above. Furthermore, by performing
the following change of variables
\begin{equation}
d_k^{\pm}=\frac{1}{\sqrt{2}}(d_k^1\pm i d^2_k)
\end{equation}
we see that the integral over $d_k^-$ can be easily performed and
we are left with the following expression (once we express
$d^+_k,\bar{d}^+_k$ in polar coordinates and integrate over the
angular variable)
\begin{eqnarray}
\label{quantumprofile} f_5=1+\frac{Q_5}{4\pi^2} \int d^4u
\frac{e^{i \mathbf{u}.\mathbf{x}}}{|\mathbf{u}|^2}
e^{\frac{\mu^2}{4k}(u_1^2+u_2^2)} \int_0^\infty d\rho
\frac{\rho^{2 N/k+1}}{(N/k)!} e^{-\rho^2}J_0(\mu
\frac{1}{\sqrt{k}}\sqrt{u_1^2+u_2^2}\rho).
\end{eqnarray}
The integral over $\rho$ can be done explicitly (see equations
(\ref{ea2}) and (\ref{ea3})) and we are left with
\begin{eqnarray}
f_5^{quantum}=1+L_{N/k} \left( \frac{a^2}{4 N/k}\left(
\partial_1^2+\partial_2^2 \right) \right)
\frac{Q_5}{|\mathbf{x}|^2}
\end{eqnarray}
with $L_n$ the Laguerre polynomial of order $n$. Notice that,
besides the approximation of ignoring the $\delta$ function
(\ref{delta}) in the distribution, this result is exact in $N/k$.
In order to relate both results recall that
$$L_n(x)=\sum_{m=0}^n \frac{(-1)^m n!}{(n-m)!(m!)^2}x^m$$
which allows to find the following expansion for large values of
$N/k$
\begin{equation} \label{laguerrebessel} L_{N/k}(\frac{a^2 \rho^2}{4
N/k})=J_0(a\rho)-\frac{1}{N/k}\frac{a^2
\rho^2}{4}J_2(a\rho)+...
\end{equation}
From this we see explicitly that in the limit $N/k \gg 1$ the
quantum geometry coincides with the classical one. More precisely,
around asymptotic infinity the harmonic functions can be written
as a series expansion in $a^2/r^2$. If we focus on a given term
$a^{2p}/r^{2p}$ for some fixed (but arbitrarily large) $p$ then
the coefficient of such term tends to the classical coefficient as
$N/k$ tends to infinity. Note, however, that for finite $N/k$ the
quantum harmonic function is a finite order polynomial in
$a^2/r^2$ (of degree $N/k$ ) which contains a large number of
terms that are singular at the origin (and that will re-sum only
in the strict $N/k$ infinite limit). These divergences at $r=0$
may sound like a disaster, but they are actually unphysical and
due to the fact that we ignored the delta function (\ref{delta})
in the distribution (\ref{qcircprofile}). Including the delta
function will impose a cutoff on the $\rho$ integral in
(\ref{quantumprofile}), and since all singular terms are due to
the large $\rho$ behavior of the integrand in
(\ref{quantumprofile}) the cutoff will remove the singularities in
$f_5$.

From this discussion it is clear that we can trust our
semi-classical computation provided $N/k$ is large and we do not
look at the deep interior of the solution.

As for the case of the classical curve, it is instructive to
compute $\Box f_5$ for this case
\begin{eqnarray}
\Box f_5 =-4 \pi^2 Q_5 \delta(x_3) \delta(x_4) A(x_1,x_2)\\
A(x_1,x_2)=\int_0^\infty d\rho \rho J_0(\sqrt{x_1^2+x_2^2} \rho)
L_{N/k} \left(\frac{a^2 \rho^2}{4N/k}\right)
\end{eqnarray}
Until here we have not used any approximation. Using identity
(\ref{ea3}) in the appendix and approximating $\exp(\frac{a^2
\rho^2}{4N/k}) \approx 1$ one obtains
\begin{equation}
A(x_1,x_2)=\frac{e^{-N/k~r^2/a^2} \left(N/k~r^2/a^2
\right)^{N/k}}{(N/k-1)! a^2}
\end{equation}
with $r^2=x_1^2+x_2^2$. In the limit $N/k \rightarrow \infty$
$A(x_1,x_2)$ approaches $\frac{\delta(r^2/a^2-1)}{a^2}$ and the
classical and quantum results agree. For large $N/k$ $A(x_1,x_2)$
is  approximately a gaussian around $r^2 \approx a^2$ and width
$1/\sqrt{N/k}$, indeed, using Stirling's formula
\begin{equation}
A(x_1,x_2)\approx \frac{\sqrt{N/k}}{\sqrt{2\pi}}
e^{-N/k(r^2/a^2-1)}(r^2/a^2)^{N/k}
\end{equation}
So the quantum geometry corresponds to a solution of the equations
of motion in presence of smeared sources. The width of the smeared
source goes to zero in the limit $N/k\rightarrow \infty$, as
expected.

\subsection{Dipole operator}

To each supergravity solution we can associate a "dipole operator"
defined by
\begin{equation} \label{dip}
D_{sugra}=\int_0^L |\mathbf{F}(s)|^2 ds.
\end{equation}
This operator simply measures the average spread in the $\mathbb
R^4$ plane of the curve. It will become momentarily clear why we
call this a dipole operator. It is instructive to compute
(\ref{dip}) for a curve dual to a generic CFT state $|\psi
\rangle= \prod_{k=1}^\infty(c^{i
\dagger}_{k})^{N_{k_i}}|0\rangle$, {\it i.e.}:
\begin{equation}
D_{sugra}={\cal N} \int_{d,\bar{d}} \int_0^L |\mathbf{F}(s)|^2
f(d,\bar{d}) ds
\end{equation}
Using the expression for the phase space density
(\ref{genericwigner}) and performing a similar computation to the
one leading to (\ref{level}) we get
\begin{eqnarray}
D_{sugra}=\mu^2 L \left( \sum_{k} \frac{1}{k} N_{k_i} \right)
\end{eqnarray}
This happens to agree, up to normalization, with the CFT dipole
operator defined in \cite{Alday:2005xj}. There it was shown that a
thermodynamic ensemble that includes this dipole operator
reproduces the thermodynamic behavior of the small black ring.

\section{A metric for a more general conical defect?}

The aim of this section is to shed some light on the claim of
\cite{Lunin:2002iz} appendix C, where it is shown that there is no
conical defect metric with arbitrary opening angles. The main
ingredient in the proof was the requirement of smoothness of the
metric. We will try here to relax this requirement by looking at
metrics obtained by coarse graining an ensemble of (possibly non-
smooth) metrics.

The starting point is the supersymmetric conical metric
\cite{Balasubramanian:2000rt,Maldacena:2000dr}
\begin{equation} \label{cdef}
\frac{ds^2}{N} = -(r^2 + \gamma^2) \frac{dt^2}{R^2} + r^2
\frac{dy^2}{R^2} + \frac{dr^2}{r^2 + \gamma^2} + d\theta^2 +
\cos^2\theta (d\psi + \gamma \frac{dy}{R})^2 + \sin^2 \theta
(d\varphi + \gamma \frac{dt}{R})^2
\end{equation}
where N is the AdS radius and $2\pi \gamma$ is the opening angle.
It is well known that every supersymmetric conical metric is
defined by its angular momentum and $N$. The metric (\ref{cdef})
is precisely identical to the metric that we would have found in
the near-horizon limit in section~\ref{sec241} if we would also
have computed the one-forms $A,B$ and evaluated
(\ref{genericsol}), see e.g. \cite{Lunin:2002iz} for a detailed
discussion. The relation between $\gamma$ and $k$ works out to be
$\gamma=1/k$. The construction in section~\ref{sec241} therefore
provides a construction of conical defect metrics with $k$
integer, but for $k$ non-integer the construction in
section~\ref{sec241} fails. The reason is that the classical curve
$\mathbf{F}(s)$ needs to satisfy $\int_0^L \mathbf{F}(s) ds=0$, as
$\mathbf{F}(s)$ does not have a zero-mode, and this is only true
if $k$ is an integer and the curve closes.

In order to try to construct a more general conical defect metric,
we first notice that according to (\ref{supp}), the source for the
metric has to be contained in a circle of radius $a$ in the
$x_1,x_2$-plane. The most general souce term satisfying these
requirements is
\begin{equation} \label{src1}
F_1 (s) = a \cos[f(s)], \;\;\;\;\;\;\; F_2 (s) = a \sin[f(s)],
\;\;\;\;\;\;\; F_3 (s) = F_4 (s) = 0
\end{equation}
where $f(s)$ is some arbitrary function which has to satisfy
\begin{equation}
  \int_0^L e^{i f(s)} ds = 0
\end{equation}
because $\mathbf{F}(s)$ does not contain a zeromode. In addition,
the source (\ref{circprofile}) is invariant under rotations in the
$x_1,x_2$-plane. To accomplish this we need to coarse grain over
all $U(1)$ rotations of (\ref{src1}). This is most easily done by
introducing polar coordinates $x_1 + i x_2 = u e^{i \varphi}, x_3
+ x_4 = v e^{i \psi}$ so that the $U(1)$ average can be expressed
as
\begin{eqnarray}
f_5  & =  & 1 + \frac{Q_5}{2\pi L} \int_0^{2\pi} d\xi
\int_0^L \frac{ds}{| u e^{i \varphi} - a e^{i f(s) + i \xi} |^2 + v^2} \nonumber \\
f_1  & = & 1 + a^2  \frac{Q_5}{2\pi L} \int_0^{2\pi}
d\xi \int_0^L \frac{ f'(s)^2 ds}{| u e^{i \varphi} - a e^{i f(s) + i \xi} |^2 + v^2}
\nonumber\\
A & = & - a \frac{Q_5}{2\pi L} \int_0^{2\pi} d\xi \int_0^L \frac{i
f'(s) e^{i f(s) + i \xi} ds}{| u e^{i \varphi} - a e^{i f(s) + i
\xi} |^2 + v^2} . \label{ints}
\end{eqnarray}
The constraint (\ref{length}) on the curve now reads
\begin{equation}
Q_1 = a^2 \frac{Q_5}{2\pi L} \int_0^{2\pi} d\xi \int_0^L  f'(s)^2
ds  = \frac{a^2 Q_5}{L} <f'^2>
\end{equation}
Here and in the following by $<g(s)>$ we simply mean \be <g(s)> =
\int_0^L g(s)\,ds.\ee It is straight forward to evaluate the
integrals in (\ref{ints}) to get
\begin{eqnarray}
   f_5 = 1 + \frac{Q_5}{h}\\
   f_1 = 1 + \frac{Q_1}{h}\\
   A = a Q_5 \frac{<f'>}{L} \frac{u^2 + v^2 + a^2 - h}{2h} d \varphi
\end{eqnarray}
with $ h^2 = (u^2 + v^2 + a^2)^2 - 4 a^2 u^2$. In order to put it
in a form which resembles the conical defect one as much as
possible, one has to make the following change of coordinates
\begin{equation}
u^2 = (r^2 + a^2) \sin^2 \theta, \;\;\;\;\; v = r \cos \theta
\end{equation}
Using these new coordinates, the various ingredients of
(\ref{genericsol}) become
\begin{eqnarray}
 f_5 & = &  \frac{Q_5}{r^2 + a^2 \cos^2 \theta} \nonumber \\
 f_1  & = & \frac{Q_1}{r^2 + a^2 \cos^2 \theta} \nonumber \\
 A & = & \alpha \frac{a \sqrt{Q_1 Q_5}}{r^2 + a^2 \cos^2 \theta} \sin^2 \theta d\varphi \nonumber \\
 B & = & -\alpha \frac{a \sqrt{Q_1 Q_5}}{r^2 + a^2 \cos^2 \theta} \cos^2 \theta d\psi \nonumber \\
ds_4^2 & = & (r^2 + a^2 \cos^2 \theta)(\frac{dr^2}{r^2 + a^2} +
d\theta^2) +
r^2 \cos^2 \theta d\psi^2 + (r^2 + a^2) \sin^2 \theta d\varphi^2 \nonumber \\
\mathcal{C} & = & - \frac{ Q_5 r^2 \sin^2 \theta }{ r^2 + a^2
\cos^2 \theta} d\psi \wedge d\varphi
\end{eqnarray}
where $\alpha^2 = a^2 \frac{Q_5}{Q_1} (\frac{<f'>}{L})^2 =
\frac{1}{L} \frac{<f'>^2}{<f'^2>}$ is a constant introduced for
later convenience. Next we rescale $r$ by a factor of
$\frac{\sqrt{Q_1Q_5}}{R}$ and define $\gamma=\alpha
\frac{2\pi}{<f'>}$, and after some straightforward algebraic
manipulations we end up with
\begin{eqnarray} \label{long}
   \frac{ds^2}{\sqrt{Q_1Q_5}} & = & - \left(r^2 + \gamma^2 \right) (\frac{dt}{R})^2 + r^2 (\frac{dy}{R})^2 + \frac{dr^2}{r^2 +
   \gamma^2}\nonumber \\
   & &
+ \left( d\theta^2 + \sin^2 \theta (d\varphi - \alpha \gamma \frac{dt}{R})^2 + \cos^2 \theta ( d\psi - \alpha \gamma \frac{dy}{R})^2\right)  \nonumber \\
 & &  + \frac{(1-\alpha^2) \gamma^2 }{r^2 + \gamma^2 \cos^2 \theta}
 \left( \sin^2 \theta d\Sigma_1^2 + \cos^2 \theta d\Sigma_2^2
 \right)\nonumber
 \\
   \frac{C}{Q_5} & = &   (r^2 + \gamma^2 \cos^2 \theta) \frac{dt}{R} \wedge \frac{dy}{R}
   + \frac{\alpha^2 \gamma^2 - r^2 \sin^2 \theta }{r^2 + \gamma^2 \cos^2 \theta} d\psi \wedge d\varphi
   \nonumber \\
& & - \alpha \gamma \left( \cos^2 \theta \frac{dt}{R} \wedge d\psi
+ \sin^2 \theta \frac{dy}{R} \wedge d\varphi \right)
\end{eqnarray}
where we defined
\begin{eqnarray}
 d\Sigma_1^2 & = &  \sin^2 \theta d\varphi^2 + \left(r^2 + \gamma^2 \cos^2 \theta \right) (\frac{dt}{R})^2
   \nonumber \\
d\Sigma_2^2 & = &
  - \cos^2 \theta d\psi^2 + (r^2 + \gamma^2 \cos^2 \theta) (\frac{dy}{R})^2
\nonumber
\end{eqnarray}

This metric is a conical defect metric for $\alpha=1$, so the
question is which values of $\gamma $ are compatible with
$\alpha=1$. To analyze this,  we recast the constraints on $f(s)$
for $\alpha=1$ here
\begin{eqnarray}
  \int_0^L e^{i f(s)} ds = 0 \label{con1} \\
  \left(\int_0^L f'(s) ds \right)^2 = L \int_0^L (f'(s))^2 =
  \left(\frac{2\pi}{\gamma}\right)^2. \label{con2}
\end{eqnarray}
However, according to Schwarz's inequality,
\be
\left( \int_0^L f'(s) ds \right)^2 \leq L \int_0^L (f'(s))^2
\ee
for integrable functions $f'(s)$ with equality if and only if
$f'(s)$ is a constant. Thus, $\alpha \leq 1$ and $\alpha=1$ only
if $f'(s)={\rm const}$. Interestingly, the metric (\ref{long}) is
in general a perfectly acceptable metric, since $\alpha\leq 1$ is
precisely the condition for the absence of CTC's as one can derive
using the results in \cite{Elvang:2004ds}. If $\alpha=1$ then
$f'(s)={\rm const}$ together with (\ref{con1}) imply that
$f(s)=2\pi k s/L$ for some nonzero integer $k$, and $\gamma=1/k$.
We can therefore indeed only construct conical defect metrics with
$\gamma=1/k$ and $k$ integer. For $k$ noninteger, we find a bound
on $\alpha$
\be \alpha^2 \leq \left[\frac{1}{\gamma}\right]^2 \gamma^2
\ee
with $[x]$ the largest integer less than or equal to $x$. Indeed,
we cannot come arbitrarily close to a noninteger conical defect
metric in this way.

\section{Thermal ensembles}

In the following we consider the geometry corresponding to various
thermal ensembles of interest.

\subsection{M=0 BTZ}

\label{m0btz}

We start by considering the ensemble corresponding to the $M=0$
BTZ black hole. In principle one should consider a micro-canonical
ensemble with states of fixed level
$$ \hat{N} |\psi \rangle \equiv
\sum_k k c^{i \dag}_k c^i_k |\psi \rangle= N |\psi \rangle$$ We
will, instead, consider a canonical ensemble, since in the large
$N$ limit the difference between the two should vanish. The
corresponding thermal ensemble is characterized by the following
density matrix \footnote{We are going to ignore the $i$-index in
some equations where it does not play any role. We hope that this
will not create any confusion.}
\begin{equation}
    \rho = \sum_{N_k,\tilde{N}_k} \frac{|N_k \rangle \langle N_k| e^{-\beta \hat{N}} |\tilde{N}_k\rangle \langle\tilde{N}_k|}{\text{\bf Tr} e^{-\beta \hat{N}}}
\end{equation}
where $|N_k\rangle$ is a generic state labelled by collective
indices $N_k$
$$|N_k\rangle=\prod_k \frac{1}{\sqrt{N_k!}}(c_k^\dagger)^{N_k}|0
\rangle$$ and we have chosen a normalization so that $\langle N_k
| \tilde{N}_k \rangle=\delta_{N_k,\tilde{N}_k}$. The value of the
potential $\beta$ has to be adjusted such that $\langle \hat{N}
\rangle =N$. It is clear that
\begin{equation}
     \rho = \prod_n \rho_k,\hspace{0.3in}\rho_k=(1-e^{-k \beta})
     \sum_{n=0}^{\infty} e^{-n k \beta} |k,n \rangle
     \langle k,n |
\end{equation}
with $|k,n\rangle=\frac{1}{\sqrt{n!}}(c_k^\dagger)^{n}|0 \rangle$.
Then the full distribution will simply be the product
$f(d,\bar{d})=\prod_k f^{(k)}_{d_k,\bar{d}_k}$ with
$$f^{(k)}_{d_k,\bar{d}_k}=(1-e^{-k \beta})e^{-d_k
\bar{d}_k}\sum_{n=0}^{\infty}\frac{e^{-n k \beta}}{n!}(d_k
\bar{d}_k)^n=(1-e^{-k \beta})\exp{(-(1-e^{-k \beta})d_k
\bar{d}_k)}$$

We start by computing $f_5$, this is given by
\begin{equation}
f_5=\frac{Q_5}{4 \pi^2 L}{\cal N} \int d^4u \int_0^L dr
\int_{d,\bar{d}} f(d,\bar{d})\frac{e^{\sum_k \frac{u^2
\mu^2}{2k}}e^{i
\mathbf{u}.(\mathbf{x}-\mathbf{F}(r))}}{|\mathbf{u}|^2}
\end{equation}
The first term in the exponential is due to the fact that we want
to compute a normal ordered quantity, see the discussion around
(\ref{prop}). The $d, \bar{d}$-integrals are gaussian and can
easily be performed,
\begin{equation} \label{harmf5}f_5=\frac{Q_5}{4 \pi^2}\int d^4 u \frac{\exp
\left( -\frac{|\mathbf{u}|^2 \mu^2}{2} \sum_{k=1}^\infty
\frac{1}{k(1-e^{-k\beta})} + \frac{|\mathbf{u}|^2
\mu^2}{2}\sum_{k=1}^\infty \frac{1}{k} + i \mathbf{u}.\mathbf{x}
\right)}{|\mathbf{u}|^2 }.
\end{equation}
In the limit $\beta \ll 1$ the first sum in the exponent can be
approximated by
\begin{equation}\sum_{k=1}^\infty \frac{1}{k(1-e^{-k\beta})} =
\sum_{k}\frac{1}{k}+\frac{\pi^2}{6 \beta} +{\cal O}(\log{\beta})
\label{part}\end{equation} and inserting this in \ref{harmf5} we
see that the divergent piece drops out, as expected, and we are
left with
$$f_5=Q_5 \frac{1-e^{-\frac{3 \beta}{\pi^2 \mu^2}x^2}}{x^2}.$$
This resembles the AdS answer $f_5=Q_5/x^2$, but with
exponentially suppressed corrections that render $f_5$ finite at
$x=0$.

The computation of $f_1$ is slightly more involved but completely
analogous to that of $f_5$. It is given by
\begin{equation}
f_1=\partial_{v^i} \partial_{v^i} \left( \frac{Q_5}{4 \pi^2
L}{\cal N} \int d^4u \int_0^L dr \int_{d,\bar{d}}
f(d,\bar{d})\frac{e^{\sum_k \left( \frac{|\mathbf{u}|^2
\mu^2}{2k}-\frac{2 \pi^2 k \mu^2|\mathbf{v}|^2}{L^2} \right) }e^{i
\mathbf{u}.(\mathbf{x}-\mathbf{F}(r)) + i\mathbf{v}.\mathbf{F}'(r)
}}{|\mathbf{u}|^2} \right)\left.\right|_{v=0}.
\end{equation}
Notice that one can obtain all harmonic functions by taking
appropriate $v^i$ derivatives of the quantity between parentheses
and then putting $v^i=0$. The $d,\bar{d}$ integral is again
$$\sum_{k=1}^{\infty} \frac{k}{1-e^{-\beta k}}=\sum_{k=1}^{\infty}
k+\frac{\pi^2}{6 \beta^2}+{\cal O}(1/\beta)$$ we arrive at the
following result
$$f_1=\frac{2 \pi^4 \mu^2}{3 L^2 \beta^2} Q_5 \frac{1-e^{-\frac{3 \beta}{\pi^2
\mu^2}x^2}}{x^2}=Q_1 \frac{1-e^{-\frac{3 \beta}{\pi^2
\mu^2}x^2}}{x^2}$$ where in the last equality we have used the
correct values for $Q_1, L,\mu$ together with the value for
$\beta$ to be found below in (\ref{valb}). Using similar
computations one can infer that
$$A^i=0.$$

As we already mentioned, $\beta$ has to be fixed in such a way
that $\langle \hat{N} \rangle=N$. Since the occupation number of a
oscillator $c_k$ is
$$\langle \hat{N}_k \rangle=Tr{\rho \hat{N}_k}=\frac{e^{-k \beta}}{1-e^{-k \beta}}$$
we find that
$$N=\langle \hat{N} \rangle=-4~\partial_\beta \sum_{k=1}^{\infty}
\log(1-e^{-\beta k})=\frac{2 \pi^2}{3}\frac{1}{\beta^2}$$ and
therefore we fix
\be \label{valb}
\beta=\frac{\pi \sqrt{2/3}}{\sqrt{N}}.
\ee
Obviously, the thermodynamic limit $N \gg 1$ corresponds to $\beta
\ll 1$.

A final comment is in order. The geometry obtained differs from
the {\it classical} $M=0$ BTZ black hole by an exponential piece.
Following \cite{Lunin:2002qf,Mathur:2005zp} we could put a
stretched horizon at the point where this exponential factor
becomes of order one, so that the metric deviates significantly
from the classical $M=0$ BTZ solution. Thus, using this criterion
we find for the radius of the stretched horizon \footnote{The same
value is obtained if we compute the average size of the curve in
$\mathbb R^4$, $r_0^2 \approx \langle |F|^2 \rangle$.}
\be \label{shor}
r_0 \approx \frac{\mu}{\beta^{1/2}}
\ee
with corresponding entropy proportional to $N^{3/4}$. This exceeds
the entropy of the mixed state from which the geometry was
obtained, the latter grows as $N^{1/2}$. We refer to the
conclusions for a further discussion of this mismatch.

\subsection{Condensate plus thermal ensemble: the small black ring}

In this section we consider a slightly more complicated example,
namely an ensemble consisting of a condensate of $J$ oscillators
of level $q$ plus a thermal ensemble of effective level $N-q J$.
As argued in \cite{{Bena:2004tk},Alday:2005xj} such an ensemble
should describe (in a certain region of parameter space) a small
black ring of angular momentum $J$ and dipole (or Kaluza-Klein)
charge $q$.

Using the techniques developed in the previous sections we can
compute the generating harmonic function for this case as well and
we find
\begin{equation}
f_v=Q_5 L_J \left( \frac{\mu^2}{4 q} \left[ \left( \frac{2 \pi
q}{L}v_2+i
\partial_1 \right)^2+\left( \frac{2 \pi q}{L}v_1-i
\partial_2 \right)^2 \right] \right) e^{-\frac{\mu^2 \pi^2 |v|^2}{2L^2}(N-q
J)}\frac{1-e^{-\frac{2 |\mathbf{x}|^2}{\mu^2 D}}}{|\mathbf{x}|^2}
\end{equation}
where $D\approx \pi \sqrt{2/3}(N-q J)^{1/2}$ so that the geometry
is purely expressed in terms of the macroscopic quantities $N,J$
and $q$.

We would like to make contact between this geometry and the
geometry corresponding to small black rings studied in
\cite{Alday:2005xj}. As we will see, in the limit of large quantum
numbers both geometries reproduce the same one point functions.

In order to see this, first note that the exponential factor
$e^{-\frac{2 |\mathbf{x}|^2}{\mu^2 D}}$ will not contribute (as it
vanishes faster than any power at asymptotic infinity). Secondly
notice that we can use (\ref{laguerrebessel}) in order to perform
the formal expansion
\be L_J \left( \frac{\mu^2}{4q}
\mathcal{O}\right) =J_0(\mu \sqrt{\frac{J}{q}}
\mathcal{O}^{1/2})+...
\ee
In order to estimate the validity of this approximation we can
think of ${\mathcal O}$ as being proportional to
$1/|\mathbf{x}|^2$. On the other hand $\mu \sqrt{J/q}$ can be
roughly interpreted as the radius of the black ring (see
\cite{Elvang:2004ds,{Alday:2005xj}}, where this parameter is
called $R$). Hence the approximation is valid for large values of
$J$ at a fixed distance compared to the radius of the ring, very
much in the same spirit as what happened in the case of a circular
profile.

Using (\ref{besselongeom}) and the above approximations it is then
straightforward to compute the harmonic functions
\begin{equation}
f_5 = \frac{Q_5}{r^2+\mu^2
\frac{J}{q}\cos\theta},\hspace{0.3in}f_1 = \frac{Q_1}{r^2+\mu^2
\frac{J}{q}\cos\theta}
\end{equation}
where we have used a coordinate system analogous to the one used
in the discussion of the circular profile (see (\ref{delcoord}).
Hence in this approximation the geometry reduces exactly to that
of the small black ring studied in \cite{Alday:2005xj}.

\subsection{Generic thermal ensemble}

In the following we consider a generic thermal ensemble, where
each oscillator $c_{k^{i}}$ is occupied thermally with a
temperature $\beta_{k^i}$. We further will assume that
$\beta_{k^\pm}$ for the directions $1,2$ is equal to
$\beta_{k^\pm}$ for the directions $3,4$. Restricting to, say,
directions $1,2$ we are led to consider the following distribution
\be
f(d,\bar{d})=\exp\left( -(1-e^{-\beta_{k^+}})d_k^+ \bar{d}_k^+
-(1-e^{-\beta_{k^-}})d_k^- \bar{d}_k^-\right).
\ee
Following the same steps as for the case of the $M=0$ BTZ black
hole we obtain
\begin{eqnarray}
f_5 & = & Q_5 \frac{1-e^{-\frac{2|\mathbf{x}|^2}{\mu^2
D}}}{|\mathbf{x}|^2}\\
f_1& =& Q_1 \left( \frac{1-e^{-\frac{2|\mathbf{x}|^2}{\mu^2
D}}}{|\mathbf{x}|^2} -\frac{J^2}{4N\mu^4
D^2}e^{-\frac{2|\mathbf{x}|^2}{\mu^2 D}} \right)
\\
A & = & \frac{\mu^2  JR}{2}
\left(2\frac{e^{-\frac{2|\mathbf{x}|^2}{\mu^2 D}}}{\mu^2
D}-\frac{1-e^{-\frac{2|\mathbf{x}|^2}{\mu^2 D}}}{|\mathbf{x}|^2}
\right)(\cos^2\theta d\phi+\sin^2\theta d\psi)
\end{eqnarray}
where $(|\mathbf{x}|,\theta,\phi,\psi)$ are spherical coordinates
for $\mathbb R^4$ in terms of which the metric reads $ds^2 = dr^2
+ r^2(d\theta^2 + \cos^2\theta d\phi^2 + \sin^2 \theta d\psi^2)$.

We see that, rather surprisingly, the geometry depends only on a
few quantum numbers $N,J$ and $D$ which are given in terms of the
temperatures by
\begin{eqnarray}
N=2 \sum_k k \left( \frac{e^{-\beta_{k^+}}}{1-e^{-\beta_{k^+}}} +
\frac{e^{-\beta_{k^-}}}{1-e^{-\beta_{k^-}}} \right)\\
J=2 \sum_k \left( \frac{e^{-\beta_{k^+}}}{1-e^{-\beta_{k^+}}} -
\frac{e^{-\beta_{k^-}}}{1-e^{-\beta_{k^-}}} \right)\\
 D=2 \sum_k \frac{1}{k}
\left( \frac{e^{-\beta_{k^+}}}{1-e^{-\beta_{k^+}}} +
\frac{e^{-\beta_{k^-}}}{1-e^{-\beta_{k^-}}} \right).
\end{eqnarray}
As a result, the information carried by the geometry is much less
than that carried by the ensemble of microstates. In fact, only
$N$ and $J$ are visible at infinity while $D$ sets the size of the
``core'' of the geometry. We interpret this as a manifestation of
the no-hair theorem for black holes. The derivation in this
section assumes that the temperatures are all sufficiently large.
By tuning the temperatures, it is possible to condense one (like
in the small black ring case) or more oscillators. If this
happens, we should perform a more elaborate analysis, and we
expect that the dual geometrical description\footnote{It is not
difficult to see that the harmonic functions now will take the
form of multiple Laguerre polynomials with differential operator
arguments acting on the generating harmonic function of the $M=0$
BTZ solution.} corresponds to concentric small black rings. In
this case the configuration will depend on more quantum numbers
than just $N,J,D$, in particular we will find solutions where the
small black rings carry arbitrary dipole charge. Thus, once we try
to put hair on the small black hole by tuning chemical potentials
appropriately, we instead find a phase transition to a
configuration of concentric small black rings, each of which still
is characterized by just a few quantum numbers.

\section{Conclusions}

In this paper we proposed and studied a map from 1/2-BPS pure and
mixed states in the D1-D5 system to ten-dimensional geometries
that become 1/2-BPS solutions to the type IIB supergravity
equations of motion in the classical limit. We restricted our
attention to states associated to the bosonic fluctuations of the
D1-D5 system in the four transverse non-compact directions,
following the work of \cite{Lunin:2001jy}. To construct the map we
took advantage of the results of \cite{jevicki,Rychkov:2005ji}
where the symplectic form on the appropriate space of 1/2-BPS
solutions were obtained. We also crucially used the idea that
coherent states should be the most classical ones. An important
subtlety that we ran into is that we should work with the phase
space of curves in $\mathbb R^4$ of fixed length proportional to
$N$. This can be taken into account by inserting explicit
delta-functions in the phase space densities that we found. It
turns out that it is technically very difficult to work with this
delta-function, so instead we decided to work with a canonical
ensemble where curves of length $\ell$ are weighted with weight
$\exp(-\beta \ell)$ and $\beta$ is chosen in such a way that the
expectation value of $\ell$ is $N$. At large $N$ both methods
should agree, but there are important differences at finite $N$.
When we studied the simplest example of a circular curve, we found
using the canonical ensemble\footnote{More precisely, we used a
microcanonical approach to associate a state to a classical curve
and then a canonical approach to associate a phase space density
to it. If we would have used a canonical approach throughout we
would have found a simple exponential phase space density which
does not yield any singularities.} a metric which is very singular
in the interior, but we could qualitatively argue that these
singularities will disappear once we properly work in the
microcanonical ensemble. It would be interesting to study this in
more detail. We also saw that the quantum answer indeed
corresponds to a small approximately gaussian smearing of the
classical curve, whose width vanishes as $N\rightarrow \infty$.

In section~3 we have elaborated somewhat on the claim of
\cite{Lunin:2002iz} that one cannot have a conical defect metric
with arbitrary opening angle. We studied ensembles of smooth
metrics that resemble as much as possible conical defect metrics.
It turns out that one can indeed only construct conical defect
metrics with deficit angle $2\pi/k$ with $k$ integer, and
explicitly gave metrics that come closest to conical defect
metrics with other opening angles. Though we did not go out of our
way to prove that our construction is the most general one, it is
hard to see how one could avoid our conclusion.

In section~4 we studied different thermodynamic systems and their
geometric description. This was in fact the main motivation for
this work. We first looked at the $M=0$ BTZ ensemble, and found a
``quantum'' metric that is exponentially close to the classical
one. There are no corrections to the metric that scale as an
inverse power of the radius of AdS, and therefore all one-point
functions (except the one giving the total mass) vanish, in
agreement with general expectations. Interestingly, we found that
the natural place for the stretched horizon is at $r_0\approx
\mu\beta^{-1/2}$, with corresponding entropy $\sim N^{3/4}$. This
is is different from the results in
\cite{Lunin:2001jy,Lunin:2002bj,Lunin:2002qf,Mathur:2005zp} where
a stretched horizon was found which does yield the correct entropy
of order $N^{1/2}$. This stretched horizon was found using the
average wave number with a suitable occupation number, but first
computing the average wave number and then the average spread of
the curve is not the same as computing the average spread
directly. We were therefore unable to find a natural
interpretation for this stretched horizon in our approach. We
tried a few other possible definitions of the stretched horizon,
such as the average size of the curve, or by looking at curvature
invariants of the metric, but in all cases we found the same
result. It is possible that there exists a better definition, for
example one based on the absorption cross section, which does
yield the right entropy, and this is an interesting subject for
further study which may affect other small black holes in string
theory as well. Incidentally, having a larger stretched horizon
with more entropy does not contradict any law of physics, but a
smaller one would, since that would violate the Bekenstein-Hawking
bound. A larger one could simply mean that there are many other
microstates with less supersymmetry whose geometry also fits
inside this particular stretched horizon.

Another example we studied was the small black ring. We managed to
reproduce the geometry and one-point functions discussed in
\cite{Alday:2005xj}, thus providing further evidence that the
microscopic picture of \cite{Bena:2004tk} is the correct one.
Again, there is a stretched horizon whose associated entropy
$(N-qJ)^{3/4}$ deviates from the expected result $(N-qJ)^{1/2}$
just like the $M=0$ BTZ case.

Finally we considered more general ensembles with different
temperatures for each oscillator species. It turns out that the
metric is characterized by only three quantum numbers, $N,J$ and
$D$, where $N$ and $J$ are the energy and angular momentum as seen
at infinity, and $D$ is related to the size of the core of the
solution, i.e. to the stretched horizon. This is very reminiscent
of the no-hair theorem. The situation changes once we allow some
oscillators to have become macroscopically occupied, i.e. form a
condensate just as in Bose-Einstein condensation. In this case we
expect to find a geometry corresponding to concentric small black
rings, as discussed at the end of section~4.3, and this would
clearly be worth exploring further.

There are several avenues for further study, such as exploring in
more detail the relation between CFT correlation functions and
supergravity solutions. In principle our setup predicts the
one-point functions of all operators in arbitrary half-BPS states
and ensembles, and it would be interesting to try to reproduce
this from the CFT point of view. Another direction is to extend
our approach to the $1/4$-BPS case which does admit black hole
solutions with a macroscopic horizon. Unfortunately, there is to
date no complete classification of $1/4$-BPS solutions, but
perhaps one can already make progress using the subset of
solutions that have been found so far. We leave all these issues
to future work.

\section*{Acknowledgments}

\noindent We would like to thank V. Balasubramanian, V. Jejjala,
O. Lunin, S. Mathur, S. Rychkov, M. Shigemori and J. Simon for
useful discussions. This work was supported in part by the
stichting FOM.

\appendix

\section{Useful identities}

\begin{eqnarray}
\delta(x_1^2 +x_2^2-a^2)& =& \frac{1}{2}\int_0^{2\pi}ds
\delta(x_1-a
\cos s) \delta(x_2-a \sin s)\nonumber \\
& = & \frac{1}{8 \pi^2} \int_0^{2\pi}ds \int du_1 du_2 \exp \left(
i
u_1(x_1-a \cos s)+i u_2(x_2-a \sin s)\right) \label{ea1} \\
& = &  \frac{1}{2} \int_0^{\infty} d\rho \rho J_0(a \rho)
J_0(\sqrt{x_1^2+x_2^2} \rho) \nonumber
\end{eqnarray}

\begin{equation} \label{ea2}
\int_0^\infty d\rho \frac{\rho^{2N+1}}{N!}e^{-\rho^2}J_0(A
\rho)=\frac{1}{2}{}_1F_1(1+N,1;-A^2/4)
\end{equation}

\begin{equation} \label{ea3}
L_N(x)=\frac{e^x}{N!}\int_0^\infty e^{-t} t^N J_0(2 \sqrt{tx})dt
\end{equation}



\begin{thebibliography}{99}

\bibitem{Balasubramanian:2005mg}
  V.~Balasubramanian, J.~de Boer, V.~Jejjala and J.~Simon,
  ``The library of Babel: On the origin of gravitational thermodynamics,''
  JHEP {\bf 0512} (2005) 006
  [arXiv:hep-th/0508023].

\bibitem{Lin:2004nb}
  H.~Lin, O.~Lunin and J.~M.~Maldacena,
  ``Bubbling AdS space and 1/2 BPS geometries,''
  JHEP {\bf 0410} (2004) 025
  [arXiv:hep-th/0409174].

\bibitem{e0}
  S.~Corley, A.~Jevicki and S.~Ramgoolam,
  ``Exact correlators of giant gravitons from dual N = 4 SYM theory,''
  Adv.\ Theor.\ Math.\ Phys.\  {\bf 5}, 809 (2002)
  [arXiv:hep-th/0111222].

  \bibitem{e1}
  D.~Berenstein,
  ``A toy model for the AdS/CFT correspondence,''
  JHEP {\bf 0407}, 018 (2004)
  [arXiv:hep-th/0403110].

\bibitem{Buchel:2004mc}
  A.~Buchel,
  ``Coarse-graining 1/2 BPS geometries of type IIB supergravity,''
  arXiv:hep-th/0409271.

\bibitem{Suryanarayana:2004ig}
  N.~V.~Suryanarayana,
  ``Half-BPS giants, free fermions and microstates of superstars,''
  JHEP {\bf 0601} (2006) 082
  [arXiv:hep-th/0411145].

\bibitem{Shepard:2005zc}
  P.~G.~Shepard,
  ``Black hole statistics from holography,''
  JHEP {\bf 0510} (2005) 072
  [arXiv:hep-th/0507260].

\bibitem{Silva:2005fa}
  P.~J.~Silva,
   ``Rational Foundation Of Gr In Terms Of Statistical Mechanic In The Ads/Cft
  Framework,''
  JHEP {\bf 0511} (2005) 012
  [arXiv:hep-th/0508081].

\bibitem{e2}
  S.~Giombi, M.~Kulaxizi, R.~Ricci and D.~Trancanelli,
  ``Half-BPS geometries and thermodynamics of free fermions,''
  arXiv:hep-th/0512101.

\bibitem{e3}
  A.~Dhar, G.~Mandal and M.~Smedback,
  ``From gravitons to giants,''
  JHEP {\bf 0603}, 031 (2006)
  [arXiv:hep-th/0512312].


\bibitem{e4}
  L.~Bonora, C.~Maccaferri, R.~J.~Scherer Santos and D.~D.~Tolla,
  ``Bubbling AdS and vacuum string field theory,''
  Nucl.\ Phys.\ B {\bf 749}, 338 (2006)
  [arXiv:hep-th/0602015].

  \bibitem{e5}
  V.~Balasubramanian, B.~Czech, K.~Larjo and J.~Simon,
  ``Integrability vs. information loss: A simple example,''
  arXiv:hep-th/0602263.


\bibitem{Balasubramanian:2000rt}
  V.~Balasubramanian, J.~de Boer, E.~Keski-Vakkuri and S.~F.~Ross,
   ``Supersymmetric conical defects: Towards a string theoretic description  of
  black hole formation,''
  Phys.\ Rev.\ D {\bf 64}, 064011 (2001)
  [arXiv:hep-th/0011217].

\bibitem{Maldacena:2000dr}
  J.~M.~Maldacena and L.~Maoz,
  ``De-singularization by rotation,''
  JHEP {\bf 0212}, 055 (2002)
  [arXiv:hep-th/0012025].



\bibitem{Lunin:2001jy}
  O.~Lunin and S.~D.~Mathur,
  ``AdS/CFT duality and the black hole information paradox,''
  Nucl.\ Phys.\ B {\bf 623} (2002) 342
  [arXiv:hep-th/0109154].

\bibitem{Lunin:2002qf}
  O.~Lunin and S.~D.~Mathur,
   ``Statistical interpretation of Bekenstein entropy for systems with a
  stretched horizon,''
  Phys.\ Rev.\ Lett.\  {\bf 88}, 211303 (2002)
  [arXiv:hep-th/0202072].


\bibitem{Lunin:2002bj}
  O.~Lunin, S.~D.~Mathur and A.~Saxena,
  ``What is the gravity dual of a chiral primary?,''
  Nucl.\ Phys.\ B {\bf 655} (2003) 185
  [arXiv:hep-th/0211292].



\bibitem{Lunin:2002iz}
  O.~Lunin, J.~M.~Maldacena and L.~Maoz,
  ``Gravity solutions for the D1-D5 system with angular momentum,''
  arXiv:hep-th/0212210.


\bibitem{Bena:2004tk}
  I.~Bena and P.~Kraus,
  ``Microscopic description of black rings in AdS/CFT,''
  JHEP {\bf 0412} (2004) 070
  [arXiv:hep-th/0408186].


\bibitem{Mathur:2005zp}
  S.~D.~Mathur,
  ``The fuzzball proposal for black holes: An elementary review,''
  Fortsch.\ Phys.\  {\bf 53} (2005) 793
  [arXiv:hep-th/0502050].

  \bibitem{Iizuka:2005uv}
  N.~Iizuka and M.~Shigemori,
  JHEP {\bf 0508}, 100 (2005)
  [arXiv:hep-th/0506215].

\bibitem{Taylor:2005db}
  M.~Taylor,
  ``General 2 charge geometries,''
  JHEP {\bf 0603} (2006) 009
  [arXiv:hep-th/0507223].


  \bibitem{Balasubramanian:2005qu}
  V.~Balasubramanian, P.~Kraus and M.~Shigemori,
   ``Massless black holes and black rings as effective geometries of the D1-D5
  Class.\ Quant.\ Grav.\  {\bf 22}, 4803 (2005)
  [arXiv:hep-th/0508110].

\bibitem{Alday:2005xj}
  L.~F.~Alday, J.~de Boer and I.~Messamah,
  ``What is the dual of a dipole?,''
  arXiv:hep-th/0511246.

\bibitem{jevicki}
  A.~Donos and A.~Jevicki,
  ``Dynamics of chiral primaries in AdS(3) x S**3 x T**4,''
  Phys.\ Rev.\ D {\bf 73}, 085010 (2006)
  [arXiv:hep-th/0512017].

\bibitem{Rychkov:2005ji}
  V.~S.~Rychkov,
  ``D1-D5 black hole microstate counting from supergravity,''
  JHEP {\bf 0601} (2006) 063
  [arXiv:hep-th/0512053].

\bibitem{Elvang:2004ds}
     Elvang, Henriette and Emparan, Roberto and Mateos, David
                  and Reall, Harvey S,
    "Supersymmetric black rings and three-charge supertubes",
     Phys. Rev {\bf D71} (2005)  024033
     [arXiv :hep-th/0408120]




\end{thebibliography}
\end{document}